\documentclass[reprint,amsmath,amssymb,aps,prc,
superscriptaddress,
groupedaddress,
unsortedaddress,
runinaddress,
frontmatterverbose,
]{revtex4-1}
\usepackage{graphicx}
\usepackage{dcolumn}
\usepackage{bm}
\begin{document}
\preprint{arxiv: Sushant}
\title{Role of Carrier Concentration in Swift Heavy Ion Irradiation\\ Induced Surface Modifications}
\author{Sushant Gupta$^1$$^*$, Fouran Singh$^2$, Indra Sulania$^2$ and B. Das$^1$}
\affiliation{$^1$Department of Physics, University of Lucknow,\\Lucknow-226007, India\\
$^2$Inter University Accelerator Centre (IUAC),\\New Delhi-110067, India\\
$^*$E-mail: sushant1586$@$gmail.com}
\begin{abstract}
Highly conducting $SnO_{2}$ thin films were prepared by chemical spray pyrolysis technique. One set of as-deposited films were annealed in air for 2 h at 850$^{o}$C. These as-deposited and annealed films were irradiated using $Au^{9+}$ ions with energy of 120 MeV at different fluences ranging from $1\times10^{11}$ to $3\times10^{13}$ $ions/cm^{2}$. Electrical measurement shows that as-deposited $SnO_{2}$ films are in conducting state with n = $3.164 \times 10^{20}$ $cm^{-3}$ and annealed $SnO_{2}$ films are in insulating state. The amorphized latent tracks are created only above a certain threshold value of $S_{e}$, which directly depends on the free electron concentration (n). The electronic energy loss ($S_{e}$) of 120 MeV $Au^{9+}$ ions in $SnO_{2}$ is greater than the threshold energy loss ($S_{eth}$) required for the latent track formation in annealed $SnO_{2}$ thin film, but is less than $S_{eth}$ required for as-deposited $SnO_{2}$ film. Therefore, the latent tracks are formed in the annealed $SnO_{2}$ film and not in the as-deposited $SnO_{2}$ film. Thermal spike model is used for the calculation of threshold energy loss and radius of melted zone. The possible mechanism of the structural changes and surface microstructure evolutions is briefly discussed in the light of ion's energy relaxation processes and target's conductivity. The atomic force microscopy (AFM)  study of films shows that the morphologies of irradiated films are linked with carrier concentration of target materials.
\end{abstract}
\maketitle
\section{Introduction}
Ion beams can be used in a variety of different ways to synthesize and modify materials on the nanometer scale. By adjusting ion species, ion energy,  ion fluence and irradiation geometry it is possible to tailor the ion irradiation conditions for specific needs. In ion beam irradiation process, the
kinetic energy of incident ions is transferred to atomic/lattice and electronic systems of solids through elastic collision and electronic excitation/ionization, respectively. The transferred energy to the lattice system induces atomic displacements directly. In the electronic excitation/ionization process, on the other hand, atomic displacements can be caused as a result of indirect energy transfer from irradiating ions \cite{ref1}. For ions at a higher energy ($\sim$ 100 keV/amu), the energy (a few keV/${\AA}$) is mainly deposited via electronic excitation and ionization processes
\cite{ref2,ref3}. In most insulators, when the highly localized energy deposited on target electrons is transferred from the electrons to the target lattice, an extended damage is induced along the ion path: the so-called latent track \cite{ref3,ref4}. In metals, however, atomic displacements through the electronic excitation/ionization had been considered to hardly occur because of rapid energy dissipation by a large number of free electrons \cite{ref4}. In the last two decades, nevertheless, atomic displacements induced by high-density electronic excitation have been found even in pure metallic targets \cite{ref5,ref6}. The question of the transformation of the energy deposited by the incident ion during the slowing-down process into energy stored in the target as lattice defects is still to be answered. Two mechanisms have been proposed: (i) The thermal spike (TS) model in which the kinetic energy of target electrons excited by incident ions is transferred to the lattice system through electron-lattice interaction after rapid energy diffusion in the electronic system \cite{ref7}. (ii) The Coulomb explosion (CE) model in which the electrostatic energy of the space charge created just after the ion passage is converted into coherent radial atomic movements leading to a cylindrical shock wave \cite{ref8}. Both mechanisms are sensitive to the deposited energy density as well as to the rate of energy loss (dE/dx). The TS model of latent track formation is a well established approach that has been successfully applied to insulators, semiconductors, metals, intermetallic compounds and polymers \cite{ref9,ref10,ref11,ref12,ref13,ref14,ref15,ref16,ref17}. Because of the complexity of all energy relaxation processes involved, this model is also subjected to criticism. However, the TS model seems to be the most elaborated one; furthermore, to our knowledge, currently it is the only model being able to provide at least approximate predictions on latent track formation in numerous conducting and non-conducting targets.

For swift heavy ions (SHI), i.e., ions having velocity comparable to or larger than the orbital electron velocity of the lattice atoms, the energy dissipation in the lattice takes place mainly through ionization and electronic excitation \cite{ref18,ref19,ref20,ref21}. The rapid energy transfer during the electronic excitation can result in a variety of effects in materials including amorphization, defect creation, defect annealing, crystallization, etc. There have been numerous studies on the effects of SHI irradiation in different varieties of targets including insulators, semiconductors and metals, where electronic energy loss ($[dE/dx]_{e}$) is projected as the major parameter that determines the energy relaxation processes and the resultant effects in materials \cite{ref18,ref19,ref20,ref21}. But if a systematic analysis is done on the effects of SHI irradiation in semiconductors, it can be observed that the effects produced are determined not only by the electronic energy loss ($[dE/dx]_{e}$) but also by the physical properties of the target materials \cite{ref22,ref23}. Several material properties (such as carrier concentration in the present study) that are also influenced by the synthesis conditions seem to decide the ion beam induced effects and are of extreme importance in determining the response of a material to the SHI beam \cite{ref22}. In this paper we report our studies on the effect of SHI irradiation in $SnO_{2}$ polycrystalline thin films differing in their conductivity. There exist few reports in the literature discussing the effect of SHI irradiation on the structural, optical, electrical and sensing properties of $SnO_{2}$ thin films grown using different deposition techniques \cite{ref24,ref25,ref26,ref27,ref28,ref29,ref30,ref31,ref32,ref33,ref34,ref35,ref36,ref37,ref38}. However, no systematic studies have been carried out to verify the role of material properties such as carrier concentration (n) in determining the energy relaxation processes of SHI in $SnO_{2}$ thin films.
\section{Experimental details}
Highly conducting $SnO_{2}$ thin films were prepared using chemical spray pyrolysis technique. Dehydrate stannous chloride $SnCl_{2}.$ $2H_{2}O$ (Sigma Aldrich purity $>$ 99.99\%) was used for making the spray solution for $SnO_{2}$ thin films. An amount of 11.281 gm of $SnCl_{2}.2H_{2}O$ (Sigma Aldrich purity $>$ 99.99\%) was dissolved in 5 ml of concentrated hydrochloric acid by heating at $90^{o}C$ for 15 min. The addition of HCl rendered the solution transparent, mostly, due to the breakdown of the intermediate polymer molecules. The transparent solution thus obtained and subsequently diluted by methyl alcohol, served as the precursor. The amount of spray solution was made together 50 ml. The spray solution was magnetically stirred for 1 h and finally was filtered by a syringe filter having 0.2 $\mu$m pore size before spraying on the substrate. Fused quartz slides (10 mm $\times$ 10 mm $\times$ 1.1 mm), cleaned with organic solvents, were used as substrates. During deposition, the substrate temperature was maintained at $425^{o}$C. The solution flow rate was maintained at 0.2 ml/min by the nebulizer (droplet size 0.5-10 $\mu$m). One set of as-deposited films were annealed in air for 2 h at 850$^{o}$C.

To observe the effect of SHI irradiation, these as-deposited and annealed $SnO_{2}$ thin films were irradiated with 120 MeV Au ions using a 15 MV pelletron accelerator at Inter University Accelerator Centre (IUAC), New Delhi.  Irradiation was done at six different fluences: $1\times10^{11}$, $3\times10^{11}$, $1\times10^{12}$, $3\times10^{12}$, $1\times10^{13}$ and $3\times10^{13}$ ions/$cm^{2}$. The fluence values were estimated by integrating the charges of ions falling on the samples kept inside a cylindrical electron suppressor. The ions were incident perpendicular to the surface of the samples. A high vacuum of $10^{-6}$ Torr was maintained in the target chamber during the bombardment. Ion beam was raster scanned on the film surface by a magnetic scanner for maintaining a uniform ion flux throughout the film. Beam current was kept constant during experiments and it was around 0.5 particle nanoampere. One sample of each group was left unirradiated in the chamber and was used as the reference sample. We divided our samples into two groups, viz. group A and group B. Group A consists of films which are as-deposited and irradiated (without any post deposition annealing). On the other hand, group B consists of films which underwent post deposition annealing and were further irradiated.

The gross structure and phase purity of all films were examined by X-ray diffraction (XRD) technique using a Bruker AXS, Germany X-ray diffractometer (Model D8 Advanced) operated at 40 kV and 60 mA. In the present study, XRD data of group A and group B thin films were collected in the scanning angle ($2\theta$) range $20^{o}- 60^{o}$ using $Cu-K_{\alpha}$ radiations ($\lambda$ = 1.5405 ${\AA}$). The experimental peak positions were compared with the data from the database Joint Committee on Powder Diffraction Standards (JCPDS) and Miller indices were assigned to these peaks. Atomic force microscopy (AFM) was performed with Multi Mode SPM (Digital Instrument Nanoscope IIIa) in AFM mode to examine the microstructural evolution and root mean square surface roughness of the sample before and after irradiation. Hall measurements were conducted at room temperature to estimate the film resistivity ($\rho$), donor concentration (n) and carrier mobility ($\mu$) by using the four-point van der Pauw geometry employing Keithley's Hall effect card and switching the main frame system. A specially designed Hall probe on a printed circuit board (PCB) was used to fix the sample of the size 10 mm $\times$ 10 mm. Silver paste was employed at the four contacts. The electrical resistivity and the sheet resistance of the films were also determined using the four probe method with spring-loaded and equally spaced pins. The probe was connected to a Keithley voltmeter constant-current source system and direct current and voltage were measured by slightly touching the tips of the probe on the surface of the films. Multiple reading of current and the corresponding voltage were recorded in order to get average values. Thickness of the deposited films was estimated by an Ambios surface profilometer and was approximately 500 nm.
\section{Results and discussion}
The electronic energy loss ($S_{e}$), nuclear energy loss ($S_{n}$) and maximum penetrable range ($R_{p}$) of the 120 MeV Au ions in $SnO_{2}$ are 27.19 keV $nm^{-1}$, 0.479 keV $nm^{-1}$ and 8490 nm, respectively \cite{ref39,ref40}. The variation of $S_{e}$ and $S_{n}$ with depth in $SnO_{2}$ matrix for 120 MeV Au ions is shown in Fig. 1. Near the surface of the film, $S_{e}$ exceeds $S_{n}$ by two orders of magnitude and is almost constant throughout the film thickness, as shown in the inset of Fig. 1. This reveals that the morphological and structural changes in $SnO_{2}$ thin film on irradiation by 120 MeV $Au^{9+}$ ions are almost exclusively due to electronic energy losses.

The swift heavy ion initially interacts with the atomic electrons (electronic subsystem) of the target material and transfers its energy to electrons (valence and core electrons) in a time less than $10^{-16}$ s.  The transfer of the excess heat energy of the excited electronic subsystem to the lattice subsystem may lead to an increase in the local temperature of the target material and therefore a high energy region is developed in the close vicinity of the ion path. The amount of energy transferred (locally) depends on the coupling between the electronic and atomic subsystems. This coupling is known as electron-phonon coupling (g) and it can be defined as the ability of electrons to transfer their energy to the lattice. According to Szenes \cite{ref10,ref11}, the value of g depends on the carrier concentration (n) in the target material.

To induce melting of the material or to create columnar defects in material a certain threshold value of $S_{e}$ is required and according to Szenes' \cite{ref11} ``thermal spike model'' the value of $S_{eth}$ depends on the material parameters such as density $\rho$, average specific heat C, melting temperature $T_{m}$, irradiation temperature $T_{irr}$, initial width of the thermal spike a(0), and electron-phonon coupling efficiency g, and can be determined from the following equation:
\begin{equation}\label{1}
  S_{eth} = \frac{\pi\rho C a^{2}(0) (T_{m} - T_{irr})}{g}
\end{equation}
The value of a(0), for semiconductors, depends on their energy band gap $E_{g}$, while for insulators it is almost constant. This ion induced thermal spike width a(0) in semiconductors can be best approximated by the expression $a(0) = b + c (E_{g})^{\frac{-1}{2}}$ where b and c are constants. Using an $E_{g}$ value of 3.6 eV for $SnO_{2}$, the value of a(0) was determined from the plot of $(E_{g})^{\frac{-1}{2}}$ versus a(0) as given in reference-10, and was found to be $\sim 6.2$ nm. The efficiency with which energy deposited in the electronic subsystem is subsequently transferred to the lattice is governed by the electron-phonon coupling parameter g where typically $g_{insulator}$ $>$ $g_{condutor}$. According to Szenes \cite{ref10,ref11}, the value of g depends on electron concentration (n). It is found that, for conductive materials (n $\geq$ $10^{20}$ $cm^{-3}$), the value of g is $\sim$ 0.092 \cite{ref10}, while for insulators it is $\sim$ 0.4 \cite{ref41}. In a later section, we have reported the results of electrical measurements. From these measurements we can say pristine samples of group A (as-deposited $SnO_{2}$) are in conducting state with n = $3.164 \times 10^{20}$ $cm^{-3}$ and pristine samples of group B (annealed $SnO_{2}$) are in insulating state. The value of $S_{eth}$ for as-deposited and annealed $SnO_{2}$ films  was calculated from equation (1) by substituting the values of material density ($\rho$ = 6.99 $g/cm^{3}$), average specific heat (C  = 0.3490 $J/gK$), melting temperature ($T_{m}$ = 1898 K) \cite{ref42,ref43},
irradiation temperature ($T_{irr}$ = 300 K) for $SnO_{2}$. For the as-deposited $SnO_{2}$ films the threshold value of track formation or amorphization $S_{eth}$ (estimated by using a(0) = 6.2 nm and g = 0.092 \cite{ref10}) is 31.92 keV $nm^{-1}$.
\begin{figure}
  \centering
  \includegraphics[height=6.6cm, width=8.5cm]{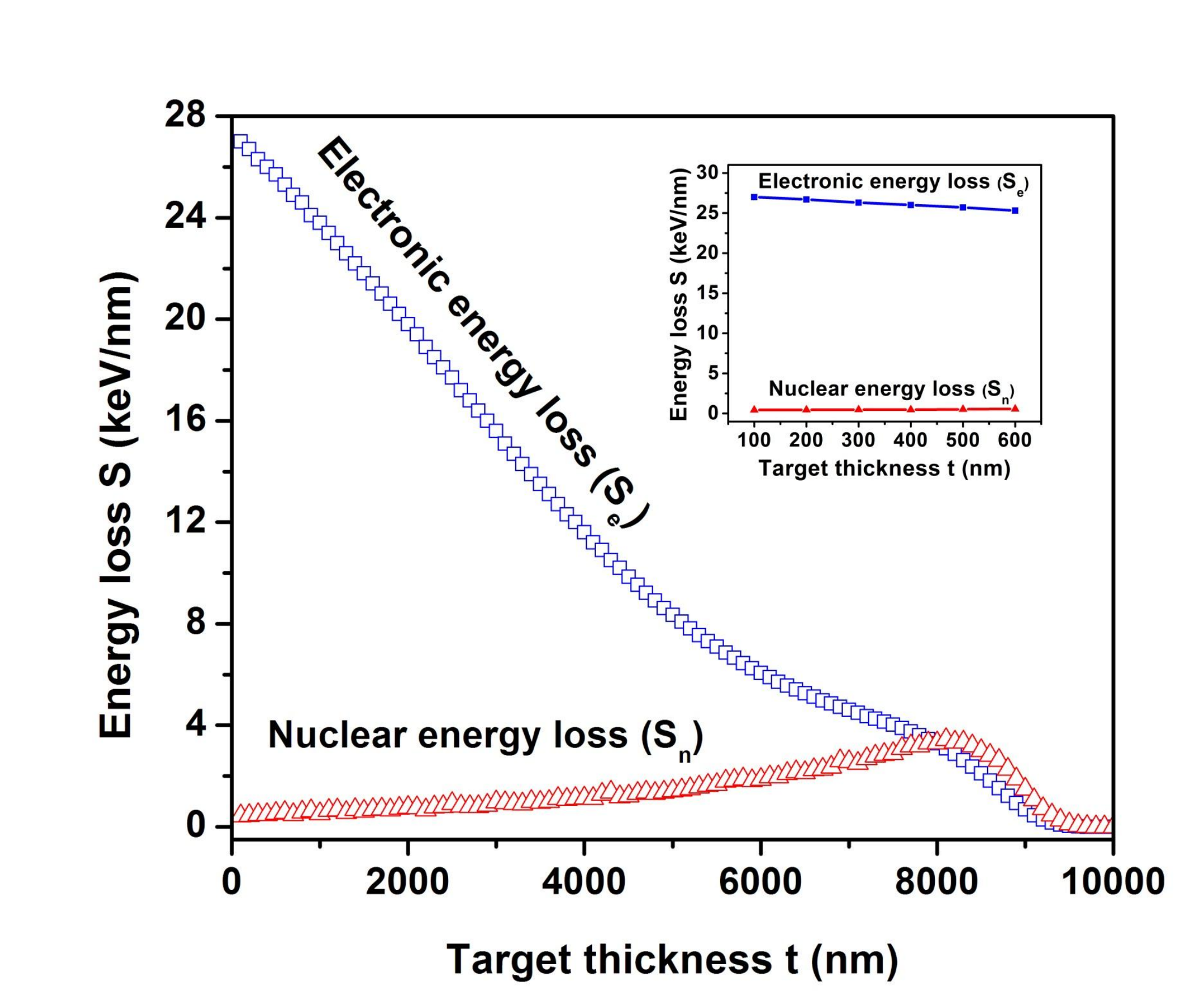}\\
  \caption{Variation of electronic energy loss $S_{e}$ and nuclear energy loss $S_{n}$ with depth for 120 MeV $Au^{9+}$ ions in $SnO_{2}$ matrix as determined by SRIM. Ratio of $S_{e}$ /$S_{n}$ is $\sim$ 60 . The inset shows the almost constant value of $S_{e}$ for even 600 nm depth from the surface of the film.}\label{1}
\end{figure}
In this case, the $S_{e}$ value is less than the threshold value required to induce amorphization/melting of material and we expect that only point defects or clusters of point defects will be produced in as-deposited $SnO_{2}$ thin films. For annealed $SnO_{2}$ films the value of $S_{eth}$ (calculated by using a(0) = 4.5 nm and g = 0.4 \cite{ref41}) is 3.868 keV $nm^{-1}$, which is much smaller than the available $S_{e}$ of 27.19 keV $nm^{-1}$ induced by 120 MeV $Au^{9+}$ ions. Therefore, latent track formation/amorphization/melting is only possible in the annealed films (samples of group B) but not in the as-deposited films (samples of group A). The value of threshold electronic energy loss $S_{eth}$ in as-deposited and annealed $SnO_{2}$ films along with values of $S_{e}$, $S_{n}$ and R are listed in the Table I. For permanent damage, the maximum temperature reached at the some local region needs to be significantly higher than the melting temperature (1898 K for $SnO_{2}$). The structure of grains is changed (spherical to ribbon/rod like) only when melting is observed and the nature/size of the new structure corresponded closely to the maximum molten region.
\begin{table}[htbp]
\renewcommand{\arraystretch}{1.5}
\caption{The values of nuclear energy loss ($S_{n}$), electronic energy loss ($S_{e}$), and range ($R_{p}$) for 120 MeV $Au^{9+}$ ions in a $SnO_{2}$ thin film estimated by SRIM \cite{ref39, ref40}. The threshold value of electronic energy loss $(S_{eth})$ and possibility of track formation are also mentioned.}
\vspace{5mm}
\centering
\begin{tabular}{||c|c|c||}
\hline
\hline
Parameters & $SnO_{2}$ & $SnO_{2}$\\
 & (as-deposited) & (annealed)\\
\hline
\hline
Nuclear energy loss & 0.479 keV $nm^{-1}$ & 0.479 keV $nm^{-1}$\\
  \hline
  Electronic energy loss & 27.19 keV $nm^{-1}$ & 27.19 keV $nm^{-1}$\\
  \hline
  Threshold energy loss & 31.92 keV $nm^{-1}$ & 3.868 keV $nm^{-1}$\\
  \hline
  Range & 8.49 $\mu$m & 8.49 $\mu$m\\
  \hline
  Track formation & Not possible & Possible \\
\hline
\hline
\end{tabular}
\end{table}

The basic assumption of thermal-spike model \cite{ref11} is that around the path of the swift heavy ion a high-temperature region is formed in the material. It is assumed that when the temperature exceeds the melting point of the material a melt is formed. Due to its small diameter, the cooling rate of the melt may reach $10^{13}$ - $10^{14}$ K/s that results in an amorphous structure when the melt solidifies \cite{ref11,ref44}. In this model \cite{ref11}, time zero is chosen when the lattice temperature on the track axis attains its maximum/peak value ($T_{p}$) and time t is measured from that moment (see Fig. 2). Let us denote by $T_{irr}$, $T_{m}$, T(r,t) the target temperature during irradiation, the melting point, and the temperature at a distance r from the ion path, respectively. If $\Delta T (r,t)$ is the local temperature increase in the thermal spike then T(r,t) = $T_{irr}$ + $\Delta T(r,t)$. This $\Delta T(r,t)$ can be described by a radial Gaussian distribution:-
\begin{equation}\label{2}
\Delta T(r,t) = \frac{Q}{\pi a^{2}(t)} e^{\frac{-r^{2}}{a^{2}(t)}}
\end{equation}
where Q = (g$S_{e}$ - L$\rho \pi R^{2}$)/($\rho C$) is determined by (partial) energy conservation and $a^{2}$(t) = $a^{2}$(0) + 4kt/($\rho$C) denotes the width of the temperature profile at later times. The quantities R = R(t) is the radius of the melted zone, $\rho$, C and L are the density, the mean specific heat, and the latent heat of melting, respectively. The parameter g determines the fraction of electronic excitation energy which is converted to heat at time zero. Here, the approximation  g$S_{e}$ $\gg$ L$\rho \pi R^{2}$ will be used. This is usually valid for materials in which $S_{e} > S_{eth}$ is observed. Interestingly, Szenes \cite{ref11} found a correlation between the threshold electronic energy loss for track formation , $S_{eth}$, and the thermal energy required to reach the melting temperature. The target material melts in the region where its temperature $(\Delta T + T_{irr})$
exceeds the melting point $T_{m}$ of material. Therefore, to reach the melting point the temperature in the thermal spike should be increased by $T_{0}$ = $T_{m}$ - $T_{irr}$.
\begin{figure}
  \centering
  \includegraphics[height=11.48cm, width=8.0cm]{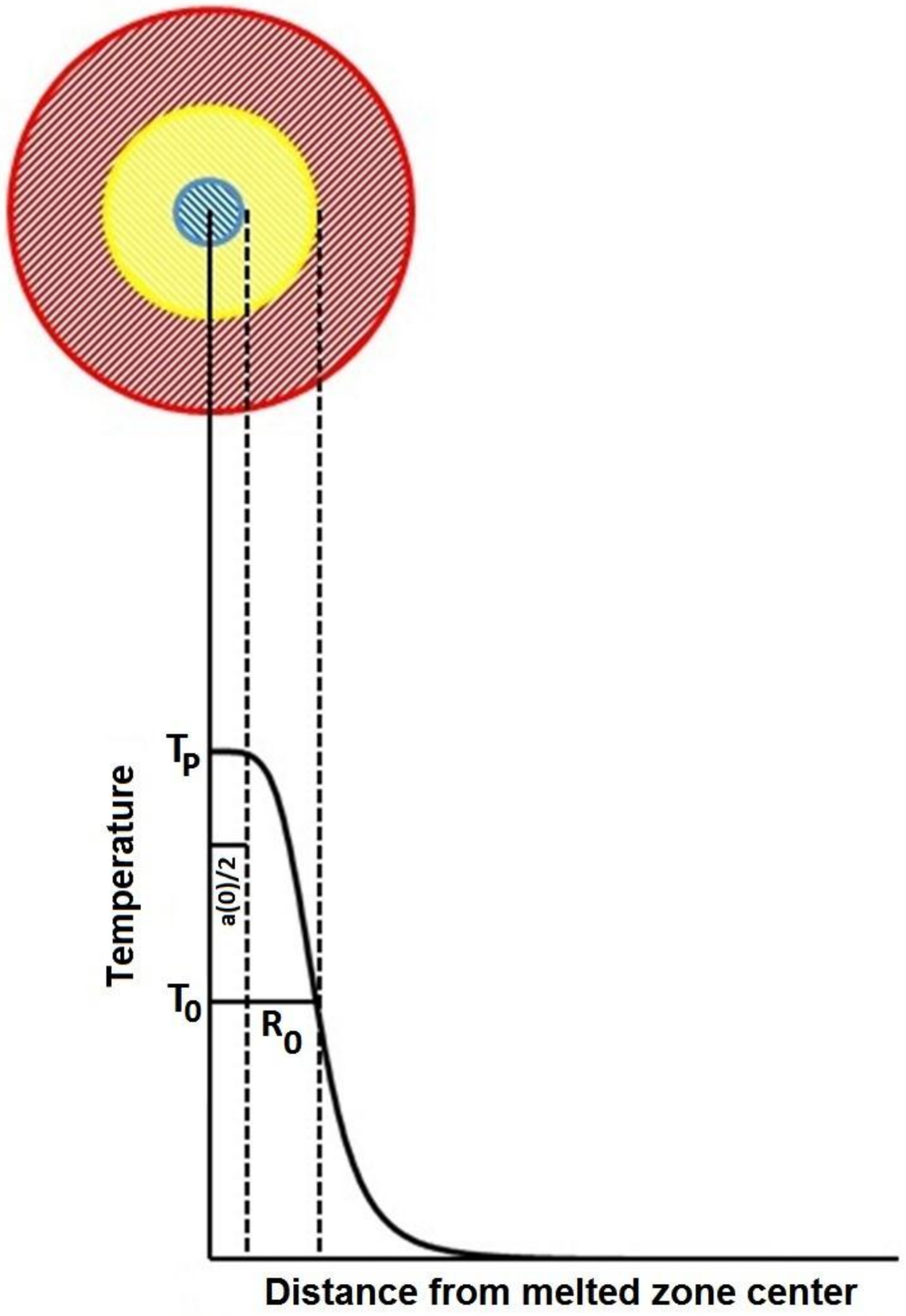}\\
  \caption{Formation of a track zone around the ion trajectory in the annealed $SnO_{2}$ film. a(0) and 2$R_{0}$ are the widths of the thermal spike and track diameter, respectively. The central blue (dark) color region shows the width of the thermal spike, a(0), and its adjacent yellow (light) color region on both sides shows the thermal-spike-induced melted zone in the annealed $SnO_{2}$ film. $T_{P}$ and $T_{0} = T_{m} - T_{irr}$ are the peak temperature and thermal spike temperature required for melting the $SnO_{2}$ film, respectively.}\label{2}
\end{figure}
The maximum value R = $R_{0}$ can be obtained from the condition dr/dt = 0 at $\Delta T$ = $T_{0}$. If the temperature at r = a(t) is denoted by $T_{a}$ then R = $R_{0}$ when $T_{a}$ = $T_{0}$.  The width of the temperature distribution a(t) increases and $T_{a}$ decreases with time in the cooling spike. If initially $T_{0}$ $>$ $T_{a}$, at t = 0 [R(0) $<$ a(0)], then $T_{0}$ = $T_{a}$ never fulfills and the melted zone will have its maximum diameter at t = 0. If $T_{0}$ $<$ $T_{a}$ at t = 0 [R(0) $>$ a(0)] then the melted zone expands up to t = $t^{'}$ when $T_{a}$ = $T_{0}$, [$R_{0}$ = R($t^{'}$) = a($t^{'}$)] and further it shrinks for t $>$ $t^{'}$ (see Fig. 2). Thus, the two simple equations for radius of melted zone (track radius) was obtained by Szenes \cite{ref11,ref13} using the condition of maxima in Eq. 2.
\begin{equation}\label{3}
  R_{0}^{2} = a^{2}(0)\ln [\frac{S_{e}}{S_{eth}}], ~~~ 1 \leq \frac{S_{e}}{S_{eth}} \leq 2.7
\end{equation}
\begin{equation}\label{4}
  R_{0}^{2} = [\frac{a^{2}(0)}{2.7}]\frac{S_{e}}{S_{eth}}, ~~~ \frac{S_{e}}{S_{eth}} > 2.7
\end{equation}
\begin{figure*}
\center
  \includegraphics[height=10.5cm, width=15.5cm]{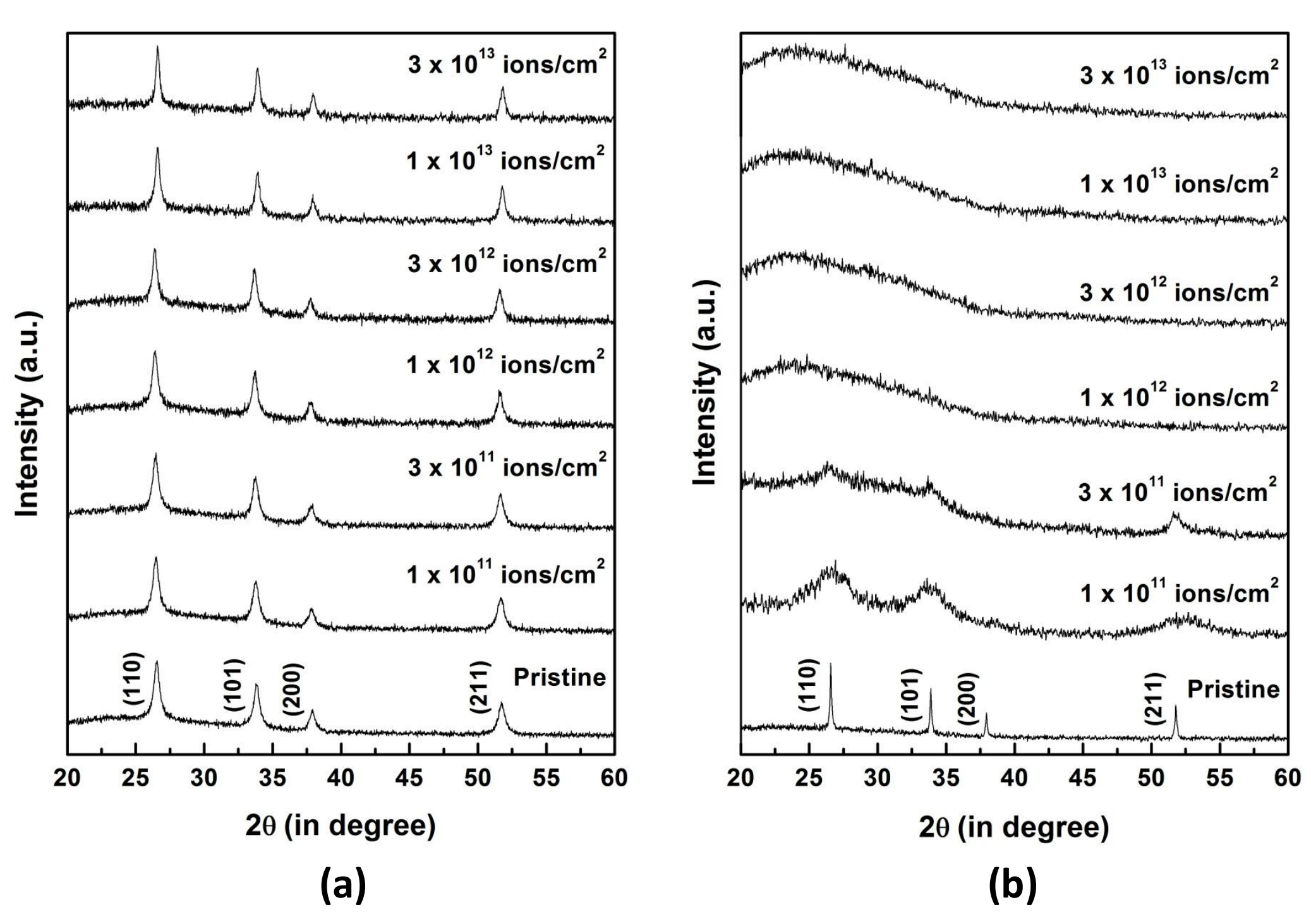}\\
  \caption{X-ray diffraction patterns of (a) Group A and (b) Group B thin films.}\label{3}
\end{figure*}
According to Eq. 3 there is a threshold electronic energy loss $S_{eth}$ below which no melting/amorphization is predicted and at this point $T_{p}$ = $T_{0}$. Expression 3 describes a logarithmic variation of the melted zone cross section A = $\pi R_{0}^{2}$  in the $1 < \frac{S_{e}}{S_{eth}} < 2.7$ range. Expression 4 is an equation of a straight line for $S_{e}$/$S_{eth}$ $>$ 2.7 going through the origin. There is a smooth transition between the linear and logarithmic regimes. At $S_{e}$ = 2.7$S_{eth}$, both expressions provide the same value for track radius.

The value of $S_{e}$/$S_{eth}$ ratio for annealed $SnO_{2}$ samples is $\sim$ 7.0. The approximate value of radius of melted zone ($R_{0}$) in annealed $SnO_{2}$ samples has been calculated from Eq. 4 as 7.26 nm. Since the estimated melted zone diameter is $\sim$ 14.52 nm, the 1 $cm^{2}$ sample will be fully covered with melted zones/ion tracks at a fluence of about $6\times10^{11}$ $ions/cm^{2}$, corresponding to 1/$\pi R_{0}^{2}$. Thus, beyond this critical fluence ($>$ $6\times10^{11}$ $ions/cm^{2}$) there will be a considerable overlapping of melted zones.
\begin{table*}[htbp]
\centering
\renewcommand{\arraystretch}{1.5}
\caption{Electrical parameters for all the samples.}
\vspace{4mm}
\begin{tabular}{||c|c|c|c|c|c|c|c||}
\hline
\hline
\centering
&\multicolumn{6}{c|}{Group A}& Group B\\ \cline{2-7} \cline{7-8}
 & Sheet& & Carrier & Degeneracy& \multicolumn{2}{c|}{Carrier mobility}& Sheet\\
 &resistance & Resistivity & concentration& temperature&\multicolumn{2}{c|}{$\mu$ ($cm^{2}V^{-1}s^{-1}$)}& resistance\\ \cline{6-7}
 Samples & $R_{s}$ & $\rho$ ($\Omega$ cm) & n ($cm^{-3}$)& $T_{D}$ (K)&Observed & Calculated &$R_{s}$\\ \hline \hline
Pristine& 35 $\Omega/\square$& 1.75$\times10^{-3}$& $3.164\times10^{20}$& 6360 & 11.291 & 9.747&$>$10 $M\Omega/\square$\\
\hline
$1\times 10^{11}$ $ions/cm^{2}$& 1.07 $k\Omega/\square$& 5.47$\times10^{-2}$& $1.357\times10^{19}$& 779 & 8.423 & -&$>$10 $M\Omega/\square$\\
\hline
$3\times 10^{11}$ $ions/cm^{2}$& 9.49 $k\Omega/\square$& 0.498& $2.381\times10^{18}$& 244 & 5.274 & -&$>$10 $M\Omega/\square$\\
\hline
$1\times 10^{12}$ $ions/cm^{2}$& 0.483 $M\Omega/\square$& 24.4& $5.157\times10^{16}$& 19 & 4.965 & -&$>$10 $M\Omega/\square$\\
\hline
$3\times 10^{12}$ $ions/cm^{2}$& 3.17 $M\Omega/\square$& 152& $8.692\times10^{15}$& 6 & 4.728 & -&$>$10 $M\Omega/\square$\\
\hline
$1\times 10^{13}$ $ions/cm^{2}$& $>$10 $M\Omega/\square$& -& -& - & - & -&$>$10 $M\Omega/\square$\\
\hline
$3\times 10^{13}$ $ions/cm^{2}$& $>$10 $M\Omega/\square$& -& -& - & - & -&$>$10 $M\Omega/\square$\\
\hline
\hline
\end{tabular}
\end{table*}

To understand the structural modifications induced by SHI irradiations, we have performed X-ray diffraction measurements. Fig. 3(a) shows the XRD patterns for group A samples. The group A samples turn out to be polycrystalline in nature with (110), (101), (200) and (211) planes of tetragonal rutile tin oxide. It is clear from XRD patterns of group A samples that there is no reduction in crystallinity after irradiation. The amorphization is induced in crystalline lattice only above a certain threshold value of $S_{e}$, which directly depends on the free electron concentration (n). In a later section, we have reported the results of Hall measurements. From these measurements we can say as-deposited $SnO_{2}$ samples are in conducting state with n $\geq$ $10^{20}$ $cm^{-3}$. These conduction electrons rapidly spread the energy of incident-ion throughout the material. Therefore, as-deposited samples require high beam energy to induce amorphization in crystalline lattice. But in the present case the value of $S_{e}$ ($\sim$ 27.19 keV $nm^{-1}$) is less than the threshold value ($S_{eth}$ $\sim$ 31.92 keV $nm^{-1}$) required to induce amorphization in as-deposited samples. Consequently, we except that only point defects or cluster of point defects will be produced after irradiation. Fig. 3(b) presents the XRD patterns of group B films. The pristine XRD pattern of group B reveals four prominent peaks ((110), (101), (200) and (211)) corresponding to tetragonal rutile structure of $SnO_{2}$. The growth of tin oxide nano-crystals and thus improvement in crystallinity after annealing is a common phenomenon \cite{ref45}. From Fig. 3(b), it is observed that intensity of diffraction peaks decreases for the films irradiated with the fluence of $1\times10^{11}$ and $3\times10^{11}$ $ions/cm^{2}$. The decrease of peak intensity is due to reduction in crystallinity of the annealed $SnO_{2}$ film. Further, with increase in ion fluence, the diffraction peaks completely disappear at the fluence of $1\times10^{12}$ $ions/cm^{2}$. The sample in this case is amorphized as a result of cascade quenching with SHI irradiation. The structural modification induced by SHI irradiation can be explained by total energy deposited in electronic excitations or ionizations in the films by energetic ion. The imparted energy of the incoming ions in annealed films at higher fluence may result in overlapping of tracks to cause lattice disordering inside grains. The value of threshold electronic energy loss $S_{eth}$ for annealed samples is very less than that of as-deposited samples. Therefore, 120 MeV $Au^{9+}$ ions are sufficient to induce amorphization in crystalline lattice of annealed samples.\\
The electrical properties of the films were estimated by resistivity and Hall effect measurements made at room temperature. The room temperature results are presented for all measured films in Table II. The as-deposited $SnO_{2}$ films (pristine samples of group A) show the best combination of electrical properties as follows: resistivity ($\rho$) of 1.75 $\times$ $10^{-3}$ $\Omega$ cm, carrier concentration (n) of 3.164 $\times$ $10^{20}$ $cm^{-3}$, and mobility ($\mu$) of 11.291 $cm^{2}V^{-1}s^{-1}$. In contrast, the resistivity of the thermally annealed films (pristine samples of group B) shows an insulating behavior. The lower resistivity in the as-deposited $SnO_{2}$ film may be due to the presence of substitutional hydrogen ($H_{O}^{+}$) [46, 47]. First principle calculations have provided evidence that usual suspects such as oxygen vacancy $V_{O}$ and tin interstitial $Sn_{i}$ are actually not responsible for n-type conductivity in majority of the cases [46-48]. These calculations indicate that the oxygen vacancies are a deep donor, whereas tin interstitials are too mobile to be stable at room temperature [47]. Recent first principle calculations have drawn attention on the role of donor impurities in unintentional n-type conductivity [46-52]. Hydrogen is indeed a especially ambidextrous impurity in this respect, since it is extremely difficult to detect experimentally [46-48]. By means of density functional calculations it has been shown that hydrogen can substitute on an oxygen site and has a low formation energy and act as a shallow donor [46-48]. Hydrogen is by no means the only possible shallow donor impurity in tin oxide, but it is a very likely candidate for an impurity that can be unintentionally incorporated and can explain observed unintentional n-type conductivity [48]. Several groups have reported on the incorporation of hydrogen in tin oxide and many have claimed that hydrogen substitutes for oxygen [46, 47, 53-67]. In the thermally annealed film, the transformation towards stoichiometry leads to an increase in the resistivity as expected for a metal-oxide semiconductor. Thermal annealing of as-deposited $SnO_{2}$ films was carried out beyond the migration temperature of $H_{O}^{+}$ defects ($\sim$ 900K) [46, 47, 66, 67] to confirm the role of $H_{O}^{+}$ defects on the film resistivity. Irradiating the as-deposited $SnO_{2}$ films with $Au^{9+}$ of fluence $1\times10^{11}$, $3\times10^{11}$, $1\times10^{12}$ and $3\times10^{12}$ ions/$cm^{2}$ increases the resistivity to 5.47 $\times$ $10^{-2}$ $\Omega$ cm, 0.498 $\Omega$ cm, 24.4 $\Omega$ cm and 152 $\Omega$ cm, respectively, from 1.75 $\times$ $10^{-3}$ $\Omega$ cm of the pristine film. The pristine films of group A exhibit a carrier concentration of 3.164 $\times$ $10^{20}$ $cm^{-3}$; this value is decreased to 1.357 $\times$ $10^{19}$ $cm^{-3}$, 2.381 $\times$ $10^{18}$ $cm^{-3}$, 5.157 $\times$ $10^{16}$ $cm^{-3}$ and 8.692$\times$ $10^{15}$ $cm^{-3}$ for the fluence of $1\times10^{11}$, $3\times10^{11}$, $1\times10^{12}$ and $3\times10^{12}$ ions/$cm^{2}$, respectively. On the other hand, the mobility of as-deposited $SnO_{2}$ films is decreased from the original value of 11.291 $cm^{2}V^{-1}s^{-1}$ to 4.728 $cm^{2}V^{-1}s^{-1}$ after irradiation with fluence of 3 $\times$ $10^{12}$ $ions/cm^{2}$. Above a typical fluence of 3 $\times$ $10^{12}$ $ions/cm^{2}$, electrical measurements show that the sheet resistance of irradiated film is of the order of 10 M$\Omega$/$\Box$. The swift heavy ion irradiation is a very effective technique to create point defects in the conducting target material (when $S_{e}$ $<$ $S_{eth}$). Both donor ($Sn_{O}^{4+}$) and acceptor ($V_{Sn}^{4-}$, $O_{i}^{2-}$, $O_{Sn}^{2-}$) type of native defects can be created with the help of ion beam. These defects are created in accordance with its respective formation energy; those having less formation energy are created in large number and vice-versa. SHI irradiation can also anneal-out the pre-existing defects ($H_{O}^{+}$) of as-deposited samples. This self-annealing of pre-existing defects ($H_{O}^{+}$) depends upon the ion fluence of irradiation. Almost all pre-existing defects ($H_{O}^{+}$) may anneal-out at high fluence irradiation. The variation in conductivity of irradiated samples may be caused by these mixed effects.

The temperature dependence of electrical resistivity in the range 30-200$^{o}C$ indicates that as-deposited $SnO_{2}$ films (pristine samples of group A) are degenerate semiconductors. The film degeneracy was further confirmed by evaluating degeneracy temperature $T_{D}$ of the electron gas by the expression \cite{ref68,ref69}:
\begin{equation}\label{5}
  T_{D} \simeq (\frac{\hbar^{2}}{2m^{*}k_{B}})(3\pi^{2}n)^\frac{2}{3} = E_{F}/k_{B},
\end{equation}
where $m^{*}$ is the reduce effective mass and n is the electron concentration. The degeneracy temperature of all investigated samples is clearly displayed in  Table II. It can be seen that $T_{D}$ of as-deposited $SnO_{2}$ (pristine) films are well above room temperature, at around 6000 K. Here, we have tried to identify the main scattering mechanisms that influence the mobility of as-deposited $SnO_{2}$ films. There are many scattering mechanisms such as grain-boundary scattering, domain scattering, surface scattering, interface scattering, phonon scattering (lattice vibration), neutral, and ionized impurity scattering which may influence the mobility of the films \cite{ref70,ref71}. The interaction between the scattering centres and the carriers determines the actual value of the mobility of the carriers in the samples.  In the interpretation of the mobility obtained for as-deposited $SnO_{2}$ films, one has to deal with the problem of mixed scattering of carriers. To solve this problem, one has to identify the main scattering mechanism and then determine their contributions. The $SnO_{2}$ films prepared here are polycrystalline. They are composed of grains joined together by grain boundaries, which are transitional regions between different orientations of neighboring grains. These boundaries between grains play a significant role in the scattering of charge carriers in polycrystalline thin films. The grain boundary scattering has an effect on the total mobility only if the grain size is approximately of the same order as the mean free path of the charge carriers ($D \sim \lambda$). The mean free path for the degenerate samples can be calculated from known mobility ($\mu$) and carrier concentration (n) using the following expression \cite{ref69,ref71}:
\begin{equation}\label{6}
  \lambda = (3\pi^{2})^{\frac{1}{3}}(\frac{\hbar\mu}{e})n^{\frac{1}{3}},
\end{equation}
The mean free path value calculated for the as-deposited $SnO_{2}$ film is 1.569 nm which is considerable shorter than grain size (D $\sim$ 50 nm) estimated using AFM. Moreover, the effect of crystallite interfaces is weaker in semiconductors, with n $\geq$ $10^{20}$ $cm^{-3}$, observed here, as a
consequence of the narrower depletion layer width at the interface between two grains \cite{ref72}. Based on above discussion it is concluded that grain boundary scattering is not a dominant mechanism.

The mobility of the free carrier is not affected by surface scattering unless the mean free path is comparable to the film thickness \cite{ref73}.
Mean free path value calculated for the as-deposited $SnO_{2}$ film is 1.569 nm, which is much smaller than the film thickness (t$\sim$500 nm). Hence, surface scattering can be ruled out as the primary mechanism. Scattering by acoustical phonons \cite{ref74} apparently plays a subordinate role in the as-deposited $SnO_{2}$ films because no remarkable temperature dependence have been observed between 30 and $200^{o}$C. Moreover, neutral impurity scattering can be neglected because the neutral defect concentration is negligible in the as-deposited $SnO_{2}$ films \cite{ref69,ref71}. Electron-electron scattering, as suggested to be important in Ref. 71, can also be neglected as it does not change the total electron momentum and thus not the mobility. In high crystalline $SnO_{2}$ films, scattering by dislocations and precipitation is expected to be of little importance \cite{ref75}.

\begin{table*}[htbp]
\centering
\renewcommand{\arraystretch}{1.55}
\caption{Morphological studies for all the samples. }
\vspace{3mm}
\begin{tabular}{||c|c|c|c|c|c|c||}
\hline
\hline
\centering
&\multicolumn{3}{c|}{Group A}& \multicolumn{3}{c||}{Group B}\\ \cline{2-5} \cline{4-7}
 & Particle& Particle& Rms & Particle& Particle& Rms\\
  Samples&shape & size & roughness& shape& size& roughness\\
  \hline
Pristine                       & Spherical& 15-50 nm& 5 nm& Spherical & 50-150 nm& 15 nm\\
\hline
$1\times 10^{11}$ $ions/cm^{2}$& Spherical & 15-50 nm&  10 nm&Irregular&-& 21 nm\\
\hline
$3\times 10^{11}$ $ions/cm^{2}$& Spherical & 20-50 nm& 12 nm&Irregular&-& 13 nm\\
\hline
$1\times 10^{12}$ $ions/cm^{2}$& Spherical & 20-55 nm& 15 nm&Ribbon/Rod-like& 50-150 nm thick & 8 nm\\
\hline
$3\times 10^{12}$ $ions/cm^{2}$& Spherical & 20-55 nm& 13 nm&Ribbon/Rod-like& 50-150 nm thick& 16 nm\\
\hline
$1\times 10^{13}$ $ions/cm^{2}$& Spherical & 30-60 nm& 25 nm&Ribbon/Rod-like& 50-200 nm thick& 15 nm\\
\hline
$3\times 10^{13}$ $ions/cm^{2}$& Spherical& 30-60 nm& 23 nm&Ribbon/Rod-like& 200-300 nm thick& 15 nm\\
\hline
\hline
\end{tabular}
\end{table*}

Another scattering mechanism popular in unintentionally doped semiconductors is the ionized impurity scattering. According to the Brooks-Herring formula \cite{ref76}, the relaxation time for coupling to ionized impurities is in the degenerate case, given by
\begin{equation}\label{7}
  \tau_{i} = \frac{(2m^{*})^{\frac{1}{2}}(\epsilon_{o}\epsilon_{r})^{2}(E_{F})^\frac{3}{2}}{\pi e^{4}N_{i}f(x)},
\end{equation}
with $N_{i}$ the carrier concentration of ionized impurities and f(x) given by
\begin{equation}\label{8}
  f(x) = ln(1+x) - \frac{x}{1+x},
\end{equation}
with
\begin{equation}\label{9}
  x = \frac{8m^{*}E_{F}R_{S}^{2}}{\hbar^{2}},
\end{equation}
The screening radius $R_{S}$ is given by
\begin{equation}\label{10}
  R_{S} = (\frac{\hbar}{2e})(\frac{\epsilon_{o}\epsilon_{r}}{m^{*}})^{\frac{1}{2}}(\frac{\pi}{3N_{i}})^{\frac{1}{6}},
\end{equation}
where $\epsilon_{r}$ is the relative dielectric permittivity and $m^{*}$ is the effective mass of the carriers.
The mobility ($\mu$) is defined as
\begin{equation}\label{11}
\mu = \frac{e\tau}{m^{*}},
\end{equation}
Substitution of the $\tau_{i}$ expression [Eq. (7)] in Eq. (11) yields the expression for mobility due to ionized impurities as
\begin{equation}\label{12}
  \mu_{i} = \frac{(\frac{2}{m^{*}})^{\frac{1}{2}}(\epsilon_{o}\epsilon_{r})^{2}(E_{F})^{\frac{3}{2}}}{\pi e^{3}N_{i}f(x)},
\end{equation}
Since all the $H_{O}^{+}$ defects present in the as-deposited $SnO_{2}$ films will be fully ionized at room temperature, impurity ion concentration will be equal to the free carrier concentration. Thus taking $N_{i}$ = n, $m^{*}$ = 0.31m, $\epsilon_{r}$ = 13.5 \cite{ref42} and using Eq. (5) in Eq. (12) we get
simplified form as
\begin{equation}\label{13}
  \mu _{i} = \frac{2.4232 \times 10^{-4}}{f(x)},
\end{equation}
with
\begin{equation}\label{14}
  x = 1.7942 \times 10^{-9} n^{\frac{1}{3}},
\end{equation}
The calculated mobility and measured mobility values for
as-deposited $SnO_{2}$ thin films are 9.747 and 11.291 $cm^{2}V^{-1}s^{-1}$ respectively, both are comparable to each other. This clearly indicates that the main scattering mechanism reducing the intra-grain mobility of the electrons in as-deposited $SnO_{2}$ films is the ionized impurity scattering. Ionized impurity scattering with singly ionized $H_{O}^{+}$ donors  best describes the mobility of as-deposited $SnO_{2}$ samples. This finding supports our assumption that $H_{O}^{+}$ defect is source of conductivity in as-deposited $SnO_{2}$ sample. The mobility of electrons in non-degenerate semiconductor increases with temperature and is independent of the electron concentration, whereas the mobility in a highly degenerate semiconductor is nearly independent of temperature and increases with electron concentration \cite{ref77,ref78}.
\begin{figure*}
  \centering
  \includegraphics[height=19.3cm, width=18.5cm]{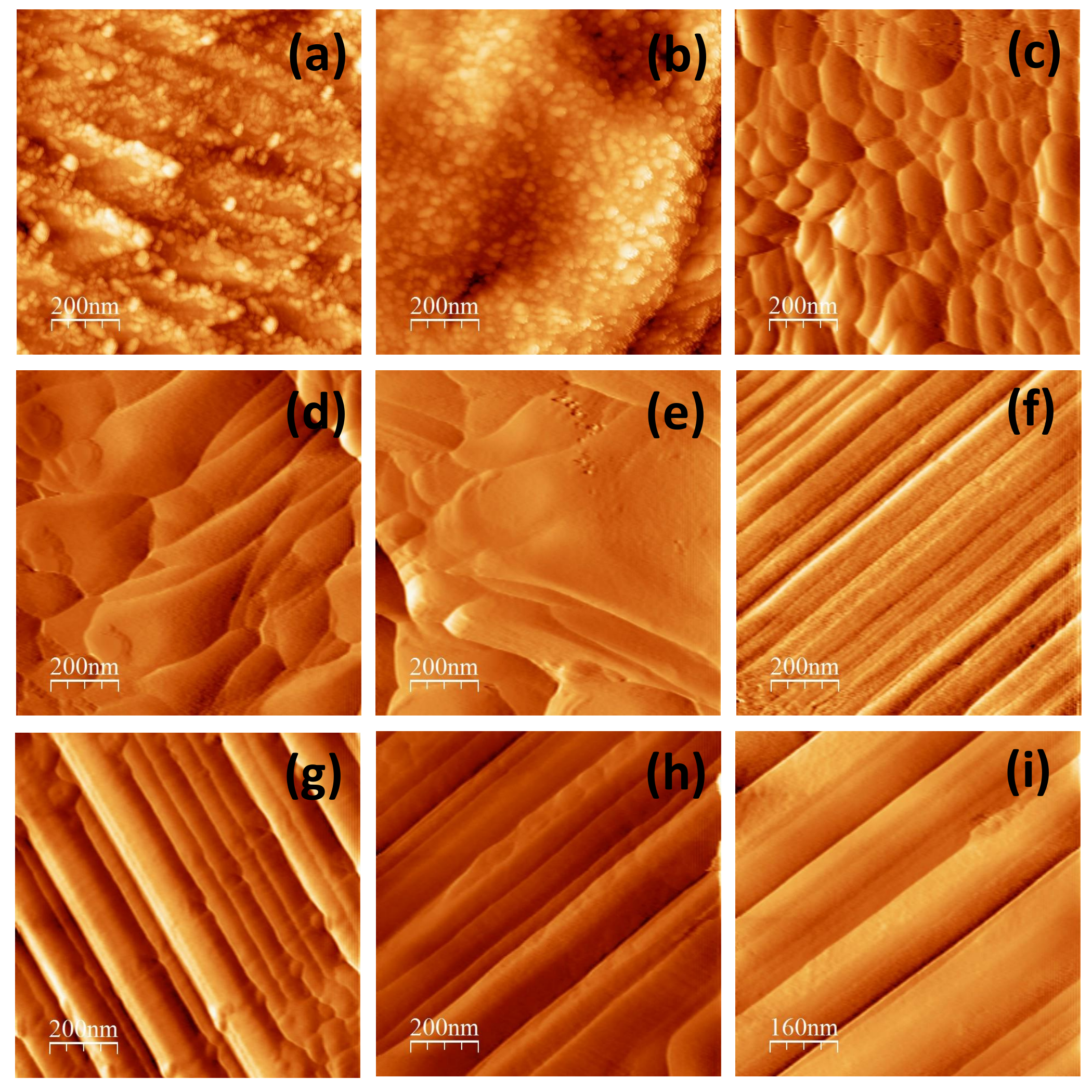}\\
  \caption{Atomic force microscopy images of $SnO_{2}$ surfaces: (a) Group A - Pristine sample, (b) Group A - $3\times10^{13}$ ions/$cm^{2}$ irradiated sample, (c) Group B - Pristine sample, (d) Group B - $1\times10^{11}$ ions/$cm^{2}$ irradiated sample, (e) Group B - $3\times10^{11}$ ions/$cm^{2}$ irradiated sample, (f) Group B - $1\times10^{12}$ ions/$cm^{2}$ irradiated sample, (g) Group B - $3\times10^{12}$ ions/$cm^{2}$ irradiated sample, (h) Group B- $1\times10^{13}$ ions/$cm^{2}$ irradiated sample, (i) Group B - $3\times10^{13}$ ions/$cm^{2}$ irradiated sample.}\label{4}
\end{figure*}
\begin{figure*}
  \centering
  \includegraphics[height=14.4cm, width=18cm]{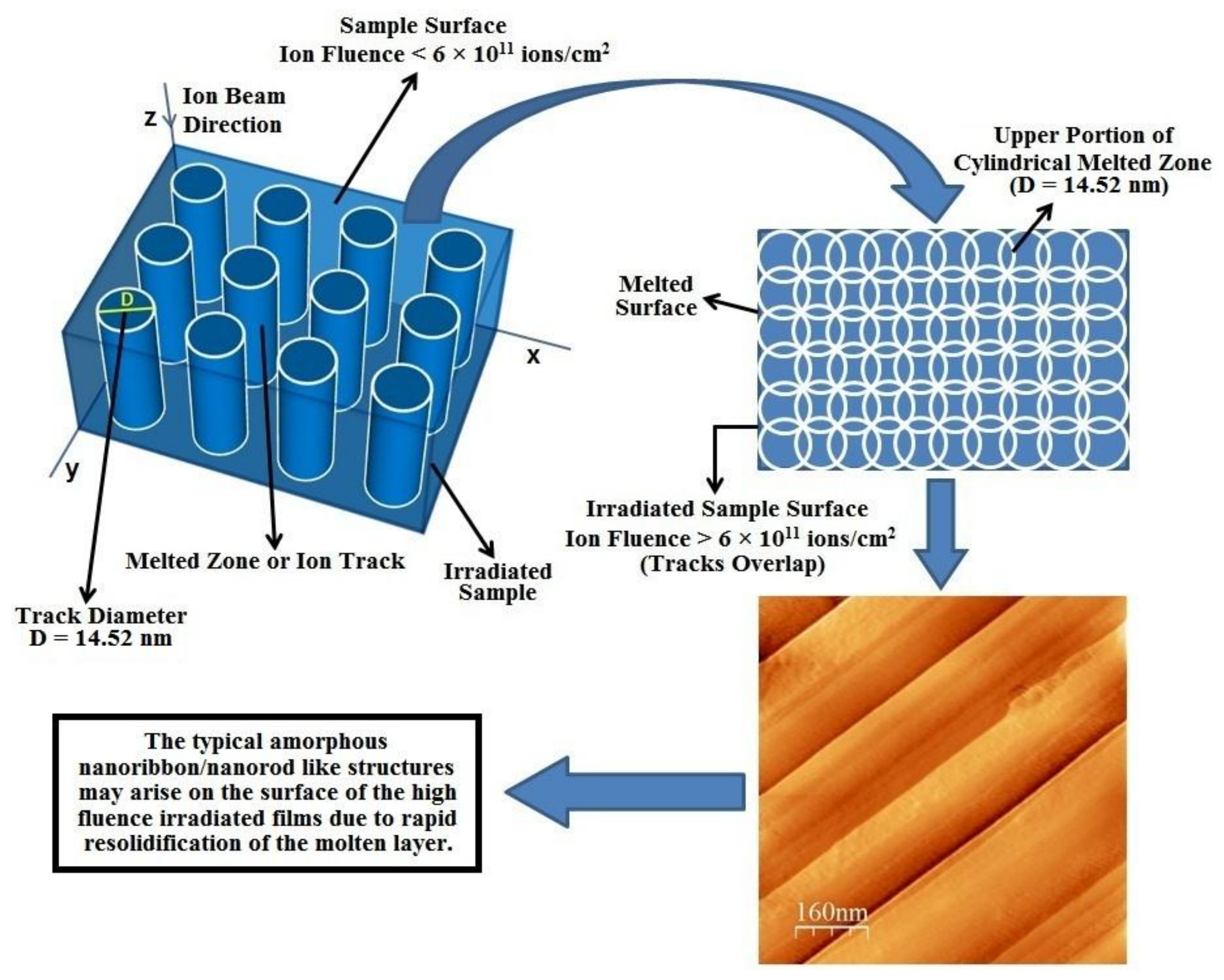}\\
  \caption{Schematic illustration of the formation mechanism of nanoribbon/nanorod like structures.}\label{5}
\end{figure*}

The surface morphology of group A and group B samples were investigated by atomic force microscopy (AFM) using a NanoScope IIIa from Digital Instruments operating in tapping mode. Fig. 4 shows the AFM micrographs of group A and group B samples. Fig. 4(a) shows the morphology of the as-deposited $SnO_{2}$ film (pristine sample of group A) which has nanostructures with broader size distribution in the range from 15 to 50 nm. The root mean square (rms) roughness is 5 nm. When this as-deposited $SnO_{2}$ film is irradiated with $3\times10^{13}$ $ions/cm^{2}$ (Fig. 4(b)), there is a significant increment in rms roughness (23 nm) and crystallite size. The dimension of nanostructures becomes 30 to 60 nm. Fig. 4(c) shows the surface morphology of an annealed $SnO_{2}$ film (pristine sample of group B). After annealing, the nanostructures become almost spherical in shape with size in the range of 50 to 150 nm and rms roughness of 15 nm. The increment in grain size is caused due to the thermal heating induced growth in crystallinity. These grains are aligned in a particular direction leading to increment in rms roughness. It is clear from the Fig. 4(c) that annealed $SnO_{2}$ films show granular nanostructure with distinct grain boundary. Figs. 4(d)-(i) shows surface morphologies of the annealed samples irradiated with $1\times10^{11}$, $3\times10^{11}$, $1\times10^{12}$, $3\times10^{12}$, $1\times10^{13}$ and $3\times10^{13}$ $ions/cm^{2}$, respectively. The irradiated samples of group B show that nanostructures get started to alteration at the lower irradiation fluence of $1\times10^{11}$ $ions/cm^{2}$. The grains are still distinct and not cobbled together at lower irradiation fluence. Remarkable changes can be seen in $3\times10^{11}$ $ions/cm^{2}$ fluence irradiated film. The nanostructures get entirely smashed via eliminating boundaries and the new aggregated grains of unusual morphology are appeared on the sample surface.  This arrangement of structures, under SHI irradiation, is expected to be governed by the confinement of a high density of energy of bombarding ions in the volume of individual grains and grain diffusion process. Fig. 4(i) shows the surface of the group B film irradiated with $3\times10^{13}$ $ions/cm^{2}$ fluence. Morphology of this film is very much different from that shown in Fig. 4(b). At this condition nanoribbon/nanorod like structures with size in the range of 200-300 nm and rms roughness of 15 nm are grown in lateral direction on the surface. The details of the morphological studies are summarized in Table-III. Morphology of irradiated film depends on the conductivity of target material. If the target material is a conductor (pristine sample of group A: $n \geq 10^{20}cm^{-3}$) then conduction electrons rapidly spread the energy of incident ion throughout the material and we find similar pristine/target sample like morphology with some dimensional change after irradiation (see Figs. 4(a) and 4(b)). But if target material is an insulator (such as pristine sample of group B) then projectile ions can not spread their energy throughout the target material and therefore they create high energy region in the close vicinity of their paths. This cylindrical region around the path of the ion may become fluid if maximum temperature reached at the centre is higher than the melting temperature of the material. The structure of grains is changed (spherical to ribbon/rod like) only when melting is observed and the nature/size of the new structure corresponded closely to the maximum molten region. The approximate value of molten zone radius on the annealed $SnO_{2}$ film surface is $\sim$ 7.26 nm. By considering 14.52 nm as the diameter of the molten zone on the annealed film surface, the critical fluence to cover the total surface with molten zones is found to be about $6 \times 10^{11}$ $ions/cm^{2}$. Thus, beyond this critical fluence there will be a considerable overlapping of molten zones (See Fig. 5). The nanoribbon/nanorod like morphology was first observed at $1 \times 10^{12}$ $ions/cm^{2}$, which is higher than the value $6 \times 10^{11}$ $ions/cm^{2}$ of ion fluence. This shows that the nanoribbon/nanorod like structures are not created by single-ion impact but rather they are the result of overlapping of molten zones. The typical amorphous nanoribbon/nanorod like structures may arise on the surface of the high fluence ($>$ $6 \times 10^{11}$ $ions/cm^{2}$) irradiated films due to rapid resolidification of the molten layer. According to the principles of solidification [79], the morphology of solidified material is controlled by the temperature gradient (G) in the liquid near the advancing interface and by the growth rate (R). Melt recoil pressure, surface tension, diffusion and evaporation dynamics may also affect the resolidification process of molten layer and consequently the structure of grains.
\section{Conclusions}
Highly conducting $SnO_{2}$ thin films were successfully deposited by spray pyrolysis technique on quartz substrates at 425$^{o}$C. One set of as-deposited films were annealed in the air by increasing the substrate temperature to 850$^{o}$C for 2h. These as-deposited and annealed films were irradiated with 120 MeV $Au^{9+}$ ions at six fluence values of $1 \times 10^{11}$, $3 \times 10^{11}$, $1 \times 10^{12}$, $3 \times 10^{12}$, $1 \times 10^{13}$ and $3 \times 10^{13}$ $ions/cm^{2}$ using 15 MV Pelletron tandem accelerator. The analysis of X-ray diffraction patterns reveals that the as-deposited and annealed tin oxide thin films are pure crystalline with tetragonal rutile phase of tin oxide which belongs to the space group P$4_{2}$/mnm (number 136). Electrical measurement shows that as-deposited $SnO_{2}$ films are in conducting state with n = $3.164 \times 10^{20}$ $cm^{-3}$ and annealed $SnO_{2}$ films are in insulating state. The results of electrical measurements suggest that $H_{O}^{+}$ defects in as-deposited $SnO_{2}$ thin films are responsible for the conductivity. Through electrical investigation it has also been found that the main scattering mechanism reducing the intra-grain mobility of the electrons in as-deposited $SnO_{2}$ films is the ionized impurity scattering. Ionized impurity scattering with singly ionized $H_{O}^{+}$ donor best describes the mobility of as-deposited $SnO_{2}$ samples. The amorphized latent tracks are created only above a certain threshold value of $S_{e}$, which directly depends on the free electron concentration (n). The electronic energy loss ($S_{e}$) of 120 MeV $Au^{9+}$ ions in $SnO_{2}$ is greater than the threshold energy loss ($S_{eth}$) required for the latent track formation in annealed $SnO_{2}$ thin film, but is less than $S_{eth}$ required for as-deposited $SnO_{2}$ film. Therefore, the latent tracks are formed in the annealed $SnO_{2}$ film and not in the as-deposited $SnO_{2}$ film. Thermal spike model is used for the calculation of threshold energy loss and radius of melted zone. The possible mechanism of the structural changes and surface microstructure evolutions is briefly discussed in the light of ion's energy relaxation processes and target's conductivity. The atomic force microscopy (AFM)  study of films reveals that the morphologies of irradiated films are linked with carrier concentration of target materials.
\begin{acknowledgments}
The authors gratefully acknowledge to D. Kanjilal, A. Tripathi, K. Asokan, P. K. Kulriya, IUAC, New Delhi and V. Ganesan, M. Gupta, M. Gangrade, UGC-DAE Consortium for Scientific Research, Indore for providing the characterization facilities. We also thank Pelletron group, IUAC, New Delhi for expert assistance in the operation of the tandem accelerator.
\end{acknowledgments}


\begin{thebibliography}{99}
\bibitem{ref1}
G. Dearnaley, Ion bombardment and implantation, Rep. Prog. Phys. 32 (1969) 405-491.
\bibitem{ref2}
N. Itoh, A. M. Stoneham, Materials Modification by Electronic Excitation, Cambridge University Press, Cambridge, 2001.
\bibitem{ref3}
N. Itoh, D. M. Duffy, S. Khakshouri, A. M. Stoneham, Making tracks: electronic excitation roles in forming swift heavy ion tracks, J. Phys.: Condens. Matter 21 (2009) 474205:1-14.
\bibitem{ref4}
D. Kanjilal, Swift heavy ion-induced modification and track formation in materials, Current Science 80 (2001) 1560-1566.
\bibitem{ref5}
A. Iwase, T. Iwata, Effect of electron excitation on radiation damage in fce metals, Nucl. Instrum. Methods Phys. Res. B 90 (1994) 322-329.
\bibitem{ref6}
A. Dunlop, D. Lesueur, P. Legrand, H. Dammak, J. Dural, Effects induced by high electronic excitations in pure metals: A detailed study in iron, Nucl. Instrum. Methods Phys. Res. B 90 (1994) 330-338.
\bibitem{ref7}
M. Toulemonde, C. Dufour, E. Paumier, Transient thermal process after a high-energy heavy-ion irradiation of amorphous metals and semiconductors, Phys. Rev. B 46 (1992) 14362-14369 .
\bibitem{ref8}
D. Lesueur, A. Dunlop, Damage creation via electronic excitations in metallic targets part II : A theoretical model, Radiat. Eff. Defects Solids 126 (1993) 163-172.
\bibitem{ref9}
G. Szenes, Ion-velocity-dependent track formation in yttrium iron garnet: A thermal-spike analysis, Phys. Rev. B 52 (1995) 6154-6157.
\bibitem{ref10}
G. Szenes, Z. E. Horvath, B. Pecz, F. Paszti, L. Toth, Tracks induced by swift heavy ions in semiconductors, Phys. Rev. B 65 (2002) 045206: 1-11.
\bibitem{ref11}
G. Szenes, General features of latent track formation in magnetic insulators irradiated with swift heavy ions, Phys. Rev. B  51 (1995) 8026-8029.
\bibitem{ref12}
G. Szenes, Information provided by a thermal spike analysis on the microscopic processes of track formation, Nucl. Instrum. Methods Phys. Res. B 191 (2002) 54-58.
\bibitem{ref13}
G. Szenes, A possible mechanism of formation of radiation defects in amorphous metals, Mater. Sci. Forum 97-99 (1992) 647-652.
\bibitem{ref14}
G. Szenes, K. Havancs$\acute{a}$k, V. Skuratov, P. Han$\acute{a}$k, L. Zsoldos, T. Ung$\acute{a}$r, Application of the thermal spike model to latent tracks induced in polymers, Nucl. Instrum. Methods Phys. Res. B 166-167 (2000) 933-937.
\bibitem{ref15}
S. Furuno, H. Otsu, K. Hojou, K. Izui, Tracks of high energy heavy ions in solids, Nucl. Instrum. Methods Phys. Res. B 107 (1996) 223-226.
\bibitem{ref16}
A. Meftah, F. Brisard, J. M. Costantini, E. Dooryhee, M. Hage-Ali, M. Hervieu, J. P. Stoquert, F. Studer, M. Toulemonde, Track formation in $SiO_{2}$ quartz and the thermal spike mechanism, Phys. Rev. B 49 (1994) 12457-12463.
\bibitem{ref17}
A. Meftah, J. M. Costantini, N. Khalfaoui, S. Boudjadar, J. P. Stoquert, F. Studer, M. Toulemonde, Experimental determination of track cross-section in $Gd_{3}Ga_{5}O_{12}$ and comparison to the inelastic thermal spike model applied to several materials, Nucl. Instrum. Methods Phys. Res. B 237 (2005) 563-574.
\bibitem{ref18}
D. K. Avasthi , G. K. Mehta, Swift Heavy Ions for Materials Engineering and Nanostructuring, Springer Series in Materials Science, Vol. 145, 2011
\bibitem{ref19}
D. K. Avasthi, J. C. Pivin, Ion beam for synthesis and modification of nanostructures, Current Science 98 (2010) 780-792.
\bibitem{ref20}
D. K. Avasthi, Modification and characterisation of materials by swift heavy ion, Defence Science Journal 59 (2009) 401-412.
\bibitem{ref21}
D. K. Avasthi, Some interesting aspects of swift heavy ions in materials science, Current Science 78 (2000) 1297-1303.
\bibitem{ref22}
Sushant Gupta, F. Singh, B. Das, Swift heavy ion irradiation-induced modifications in structural, optical, electrical and magnetic properties of Mn doped $SnO_{2-\delta}$ thin films, arXiv:1512.06119 [cond-mat.mtrl-sci] (2016) 1-46.
\bibitem{ref23}
V. V. Ison, A. Ranga Rao, V. Dutta, P. K. Kulriya, D. K. Avasthi, S. K. Tripathi, Swift heavy ion induced structural changes in CdS thin films possessing
different microstructures: A comparative study, J. Appl. Phys. 106 (2009) 023508:1-7.
\bibitem{ref24}
N. G. Deshpande, R. Sharma, Modifications in physical, optical and electrical properties of tin oxide by swift heavy $Au^{8+}$ ion bombardment, Current Applied Physics 8 (2008) 181-188.
\bibitem{ref25}
R. S. Chauhan, V. Kumar, D. C. Agarwal, D. Pratap, I. Sulania, A. Tripathi, SHI induced modifications in $SnO_{2}$ thin films: Structural, optical and surface morphological studies, Nucl. Instrum. Methods Phys. Res. B 286 (2012) 295-298.
\bibitem{ref26}
K. M. Abhirami, P. Matheswaran, B. Gokul, R. Sathyamoorthy, K. Asokan, Structural and morphological properties of Ag ion irradiated $SnO_{2}$ thin films,
IOP Conf. Ser.: Mater. Sci. Eng. 73 (2015) 012113: 1-4.
\bibitem{ref27}
F. A. Mir, K. M. Batoo, Effect of Ni and Au ion irradiations on structural and optical properties of nanocrystalline Sb-doped $SnO_{2}$ thin films, Appl. Phys. A 122 (2016) 418:1-7.
\bibitem{ref28}
T. Mohanty, P. V. Satyam, D. Kanjilal, Synthesis of nanocrystalline tin oxide thin film by swift heavy ion irradiation, J. Nanoscience and Nanotechnology 6 (2006) 2554-2559.
\bibitem{ref29}
V. Kumar, A. Jain, D. Pratap, D. C. Agarwal, I. Sulania, V. V. Siva Kumar, A. Tripathi, S. Varma, R. S. Chauhan, Modifications of nanocrystalline RF sputtered tin oxide thin film using SHI irradiation, Adv. Mat. Lett. 4 (2013) 428-432.
\bibitem{ref30}
V. Kumar, D. Pratap, A. Jain, D. C. Agarwal, I. Sulania, A. Tripathi, R. J. Chaudhary, R. S. Chauhan, Crystalline to amorphous phase transition of tin oxide nanocrystals induced by SHI at low temperature, AIP Conf. Proc. 715 (2012) 715-716.
\bibitem{ref31}
A. Sharma, K. D. Verma, M. Varshney, A. P. Singh, Y. Kumar, Growth of nanopillars in $SnO_{2}$ thin films by ion irradiation and its gas sensing properties, Adv. Sci. Lett. 4 (2011) 501-507.
\bibitem{ref32}
H. Thakur, K. K. Sharma, R. Kumar, P. Thakur, A. P. Singh, Y. Kumar, S. Gautam, K. H. Chae, On the optical properties of $Ag^{+15}$ ion beam irradiated $TiO_{2}$ and $SnO_{2}$ thin films, Journal of the Korean Physical Society 61 (2012) 1609-1614.
\bibitem{ref33}
K. M. Abhirami, P. Matheswaran, B. Gokul, R. Sathyamoorthy, K. Asokan, Swift heavy ion provoked structural, optical and electrical properties in $SnO_{2}$ thin films, Appl. Phys. A 111 (2013) 1175-1180.
\bibitem{ref34}
K. M. Abhirami, P. Matheswaran, B. Gokul, R. Sathyamoorthy, D. Kanjilal, K. Asokan, Effect of SHI irradiation on the morphology of $SnO_{2}$ thin film prepared by reactive thermal evaporation, Vacuum 90 (2013) 39-43.
\bibitem{ref35}
H. Thakur, R. Kumar, P. Thakur, N. B. Brookes, K. K. Sharma, A. P. Singh, Y. Kumar, S. Gautam. K. H. Chae, Orbital anisotropy in $SnO_{2}$ thin films and its modification by swift heavy ion irradiation, Chem. Phys. Lett. 511 (2011) 322-325.
\bibitem{ref36}
S. Rani, N. K. Puri, S. C. Roy, M. C. Bhatnagar, D. Kanjilal, Effect of swift heavy ion irradiation on structure, optical, and gas sensing properties of $SnO_{2}$ thin films, Nucl. Instrum. Methods Phys. Res. B 266 (2008) 1987-1992.
\bibitem{ref37}
S. Rani, M. C. Bhatnagar, S. C. Roy, N. K. Puri, D. Kanjilal, p-Type gas-sensing behaviour of undoped $SnO_{2}$ thin films irradiated with a high-energy ion beam, Sens. Actuators B 135 (2008) 35-39.
\bibitem{ref38}
R. C. Singh, M. P. Singh, O. Singh, P. S. Chandi, R. Kumar, Effect of 100 MeV $O^{7+}$ ions irradiation on ethanol sensing response of nanostructures of ZnO and $SnO_{2}$, Appl. Phys. A 98 (2010) 161-166.
\bibitem{ref39}
J. F. Ziegler, M. D. Ziegler, J. P. Biersack, Stopping and Ranges of Ions in Matter, SRIM Code 2008 (www.srim.org).
\bibitem{ref40}
J. F. Ziegler, J. P. Biersack, U. Littmark, The Stopping and Range of Ions in Solids, Pergamon, Vol. 1, New York, 1985.
\bibitem{ref41}
G. Szenes, F. P$\acute{a}$szti, $\grave{A}$. P$\acute{e}$ter, A. I. Popov, Tracks induced in $TeO_{2}$ by heavy ions at low velocities, Nucl. Instrum. Methods Phys. Res. B  166/167 (2000) 949-953.
\bibitem{ref42}
M. Batzill, U. Diebold, The surface and materials science of tin oxide, Prog. Surf. Sci. 79 (2005) 47-154.
\bibitem{ref43}
A. Berthelot, S. H$\acute{e}$mon, F. Gourbilleau, C. Dufour, B. Domeng$\grave{e}$s, E. Paumier, Behaviour of a nanometric $SnO_{2}$ powder under swift heavy-ion irradiation: from sputtering to splitting, Philosophical Magazine A 80 (2000) 2257-2281.
\bibitem{ref44}
M. Toulemonde, J. M. Costantini, Ch. Dufour, A. Meftah, E. Paumier, F. Studer, Track creation in $SiO_{2}$ and $BaFe_{12}O_{19}$ by swift heavy ions: A thermal spike description, Nucl. Instrum. Methods Phys. Res. B 116 (1996) 37-42.
\bibitem{ref45}
K. N. Yu, Y. Xiong, Y. Liu, C. Xiong, Microstructural change of nano-$SnO_{2}$ grain assemblages with the annealing temperature, Phys. Rev. B 55 (1997) 2666-2671.
\bibitem{ref46}
M. D. McCluskey, M. C. Tarun, S. T. Teklemichael, Hydrogen in oxide semiconductors, J. Mater. Res. 27 (2012) 2190-2198.
\bibitem{ref47}
A. K. Singh, A. Janotti, M. Scheffler, C. G. Van de Walle, Sources of electrical conductivity in $SnO_{2}$, Phys. Rev. Lett.  101 (2008) 055502:1-4.
\bibitem{ref48}
A. Janotti, J. B. Varley, J. L. Lyons, C. G. Van de Walle, Controlling the conductivity in oxide semiconductors, In: J. Wu et. al. (eds.), Functional Metal Oxide Nanostructures, Springer Series in Materials Science, 2012.
\bibitem{ref49}
A. Janotti, C. G. Van de Walle, Native point defects in ZnO, Phys. Rev. B 76 (2007) 165202-165224.
\bibitem{ref50}
A. Janotti, C. G Van de Walle, Fundamentals of zinc oxide as a semiconductor, Rep. Prog. Phys. 72 (2009) 126501:1-29.
\bibitem{ref51}
A. Janotti, C. G. Van de Walle, Oxygen vacancies in ZnO, Appl. Phys. Lett.  87 (2005) 122102:1-3.
\bibitem{ref52}
A. Janotti, C. G. Van de Walle, New insights into the role of native point defects in ZnO, J. Cryst. Growth 287 (2006) 58-65.
\bibitem{ref53}
\c{C}. Kili\c{c}, A. Zunger, n-type doping of oxides by hydrogen, Appl. Phys. Lett. 81 (2002) 73-75.
\bibitem{ref54}
W. M. Hlaing Oo, S. Tabatabaei, M. D. McCluskey, J. B. Varley, A. Janotti, C. G. Van de Walle, Hydrogen donors in $SnO_{2}$ studied by infrared spectroscopy and first-principles calculations, Phys. Rev. B 82 (2010) 193201:1-4.
\bibitem{ref55}
F. Bekisli, M. Stavola, W. Beall Fowler, L. Boatner, E. Spahr, G. L\"{u}pke, Hydrogen impurities and shallow donors in $SnO_{2}$ studied by infrared spectroscopy, Phys. Rev. B  84 (2011) 035213:1-8.
\bibitem{ref56}
J. B. Varley, A. Janotti, A. K. Singh, C.G. Van de Walle, Hydrogen interactions with acceptor impurities in $SnO_{2}$: first-principles calculations, Phys. Rev. B 79 (2009) 245206.
\bibitem{ref57}
P. D. C. King, R. L. Lichti, Y. G. Celebi, J. M. Gil, R. C. Vil$\tilde{a}$o, H. V. Alberto, J. Piroto Duarte, D. J. Payne, R. G. Egdell, I. McKenzie, C. F. McConville, S. F. J. Cox, T. D. Veal, Shallow donor state of hydrogen in $In_{2}O_{3}$ and $SnO_{2}$: implications for conductivity in transparent conducting oxides,  Phys. Rev. B 80 (2009) 081201:1-4.
\bibitem{ref58}
K. Xiong, J. Robertson, S. J. Clark, Behavior of hydrogen in wide band gap oxides, J. Appl. Phys. 102 (2007) 083710:1-13.
\bibitem{ref59}
C. G. Van de Walle, Hydrogen as a shallow center in semiconductors and oxides, Phys. Status Solidi B 235 (2003) 89-95.
\bibitem{ref60}
A. Janotti, C. G. Van de Walle, Hydrogen multicentre bonds, Nature Materials  6 (2007) 44-47.
\bibitem{ref61}
S. F. J. Cox, E. A. Davis, S. P. Cottrell, P. J. C. King, J. S. Lord, J. M. Gil, H. V. Alberto, R. C. Vil$\tilde{a}$o, J. Piroto Duarte, N. Ayres de Campos, A. Weidinger, R. L. Lichti, S. J. C. Irvine, Experimental confirmation of the predicted shallow donor hydrogen state in zinc oxide, Phys. Rev. Lett. 86 (2001) 2601-2604.
\bibitem{ref62}
D. G. Thomas, J. J. Lander, Hydrogen as a donor in zinc oxide, J. Chem. Phys. 25 (1956) 1136-1142.
\bibitem{ref63}
C. G. Van de Walle, Hydrogen as a cause of doping in zinc oxide, Phys. Rev. Lett. 85 (2000) 1012-1015.
\bibitem{ref64}
G. A. Shi, M. Saboktakin, M. Stavola, S. J. Pearton, Hidden hydrogen in as-grown ZnO, Appl. Phys. Lett. 85 (2004) 5601-5603.
\bibitem{ref65}
C. G. Van de Walle, J. Neugebauer, Universal alignment of hydrogen levels in semiconductors, insulators and solutions, Nature 423 (2003) 626-628.
\bibitem{ref66}
N. H. Nickel, Hydrogen migration in single crystal and polycrystalline zinc oxide, Phys. Rev. B  73 (2006) 195204:1-9.
\bibitem{ref67}
J. Bang, K. J. Chang, Diffusion and thermal stability of hydrogen in ZnO, Appl. Phys. Lett. 92 (2008) 132109:1-3.
\bibitem{ref68}
J. R. Bellingham, W. A. Phillips, C. J. Adkins, Electrical and optical properties of amorphous indium oxide, J. Phys.: Condens. Matter 2 (1990) 6207-6221.
\bibitem{ref69}
Sushant Gupta, B. C. Yadav, P. K. Dwivedi, B. Das, Microstructural, optical and electrical investigations of Sb-$SnO_{2}$ thin films deposited by spray pyrolysis, Materials Research Bulletin 48 (2013) 3315-3322.
\bibitem{ref70}
R. B. Hadj Tahar, T. Ban, Y. Ohya, Y. Takahashi, Tin doped indium oxide thin films: electrical properties, J. Appl. Phys. 83 (1998) 2631-2645.
\bibitem{ref71}
Sushant Gupta, The Synthesis and Characterization of Transparent Conducting Antimony Doped Tin Oxide Thin Films Deposited by Spray Pyrolysis, M.Sc. Thesis, Department of Applied Physics, Babasaheb Bhimrao Ambedkar Central University, Lucknow, India, 2012, pp. 8–18.
\bibitem{ref72}
T. Nutz, M. Haase, Wet-chemical synthesis of doped nanomaterials: optical properties of oxygen-deficient and antimony-doped colloidal $SnO_{2}$, J. Phys. Chem. B 104 (2000) 8430-8437.
\bibitem{ref73}
J. W. Orton, M. J. Powell, The Hall effect in polycrystalline and powdered semiconductors, Rep. Prog. Phys., 43 (1980) 1263-1307.
\bibitem{ref74}
V. I. Fistul and V. M. Vainshtein, Mechanism of Electron Scattering in $ln_{2}O_{3}$ Films, Sov. Phys. Solid State 8 (1967) 2769.
\bibitem{ref75}
J. G. Na, Y. R. Cho, Y. H. Kim, T. D. Lee, S. J. Park, Effects of annealing temperature on microstructure and electrical and optical properties of radio-frequency-sputtered tin-doped indium oxide films, J. Am. Ceram. Soc. 72 (1989) 698-701.
\bibitem{ref76}
V. I. Fistul, Heavily Doped Semiconductors, Plenum Press, New York, 1969, p. 86.
\bibitem{ref77}
H. K. M$\ddot{u}$ller, Electrical and optical properties of sputtered $In_{2}O_{3}$ films. I. electrical properties and intrinsic absorption, Phys. Stat. Sol. 27 (1968) 723-731.
\bibitem{ref78}
H. K. M$\ddot{u}$ller, Electrical and optical properties of sputtered $In_{2}O_{3}$ films. II. optical properties in the near infrared, Phys. Stat. Sol. 27 (1968) 733-740.
\bibitem{ref79}
M. C. Flemings, Solidification processing, London: McGraw-Hill, New York, 1974.
\end{thebibliography}
\end{document}